\documentclass[twocolumn,aps,prl,final,superscriptaddress,nobibnotes]{revtex4-1}
\usepackage[latin9]{inputenc}
\usepackage{color}
\usepackage{amsmath}
\usepackage{graphicx}
\usepackage{esint}
\usepackage[unicode=true,pdfusetitle,
 bookmarks=true,bookmarksnumbered=true,bookmarksopen=true,bookmarksopenlevel=1,
 breaklinks=true,pdfborder={0 0 1},backref=false,colorlinks=true]
 {hyperref}
\hypersetup{
 linkcolor=blue,urlcolor=blue,citecolor=blue,pdfstartview={FitH},hyperfootnotes=false}
\usepackage{breakurl}

\makeatletter
\usepackage{dcolumn}
\usepackage{bm}

\usepackage{times}

\setkeys{Gin}{width=\ifdim 0.75\Gin@nat@width>\linewidth
  \linewidth
\else
  0.75\Gin@nat@width
\fi}

\makeatother

\begin{document}

\title{Electrocaloric effects in the lead-free Ba(Zr,Ti)O$_{3}$ relaxor
ferroelectric from atomistic simulations}

\author{Zhijun Jiang}

\affiliation{School of Electronic and Information Engineering \& State Key Laboratory
for Mechanical Behavior of Materials, Xi'an Jiaotong University, Xi'an
710049, China}

\affiliation{Physics Department and Institute for Nanoscience and Engineering,
University of Arkansas, Fayetteville, Arkansas 72701, USA }

\author{Sergei Prokhorenko$^{*}$}

\affiliation{Physics Department and Institute for Nanoscience and Engineering,
University of Arkansas, Fayetteville, Arkansas 72701, USA }

\author{Sergey Prosandeev}

\affiliation{Physics Department and Institute for Nanoscience and Engineering,
University of Arkansas, Fayetteville, Arkansas 72701, USA }

\affiliation{Institute of Physics and Physics Department of Southern Federal University,
Rostov-na-Donu 344090, Russia}

\author{Y. Nahas}

\affiliation{Physics Department and Institute for Nanoscience and Engineering,
University of Arkansas, Fayetteville, Arkansas 72701, USA }

\author{D. Wang}

\affiliation{School of Electronic and Information Engineering \& State Key Laboratory
for Mechanical Behavior of Materials, Xi'an Jiaotong University, Xi'an
710049, China}

\author{J. Íñiguez}

\affiliation{Materials Research and Technology Department, Luxembourg Institute
of Science and Technology, 5 avenue des Hauts-Fourneaux, L-4362 Esch/Alzette,
Luxembourg}

\author{E. Defay}

\affiliation{Materials Research and Technology Department, Luxembourg Institute
of Science and Technology, 5 avenue des Hauts-Fourneaux, L-4362 Esch/Alzette,
Luxembourg}

\author{L. Bellaiche}

\affiliation{Physics Department and Institute for Nanoscience and Engineering,
University of Arkansas, Fayetteville, Arkansas 72701, USA }
\begin{abstract}
Atomistic effective Hamiltonian simulations are used to investigate
electrocaloric (EC) effects in the lead-free Ba(Zr$_{0.5}$Ti$_{0.5}$)O$_{3}$
(BZT) relaxor ferroelectric. We find that the EC coefficient varies
non-monotonically with the field at any temperature, presenting a
maximum that can be traced back to the behavior of BZT's polar nanoregions.
We also introduce a simple Landau-based model that reproduces the
EC behavior of BZT as a function of field and temperature, and which
is directly applicable to other compounds. Finally, we confirm that,
for low temperatures (i.e., in non-ergodic conditions), the usual
indirect approach to measure the EC response provides an estimate
that differs quantitatively from a direct evaluation of the field-induced
temperature change. 
\end{abstract}
\maketitle

\section{Introduction}

The electrocaloric (EC) effect characterizes the change in temperature
induced by a change in electric field \cite{Lines1997,Jona1993,Scott2011,Zhang2006,Zhang_APL_2006,Kutnjak2015},
with the electrocaloric coefficient being defined as $\alpha=\left.\frac{\partial T}{\partial{\cal E}}\right|_{S}$,
where $T$ is the temperature, ${\cal E}$ is the electric field and
$S$ is the entropy. It has the potential to be an efficient solid-state
refrigeration for a broad range of applications \cite{Bai2010,Moya2014,Kutnjak2015,Zhang2014}.
Numerous studies have been recently conducted via measurements, phenomenologies
and atomistic simulations (see, e.g., Refs \cite{Liu2013,Liu2014,Sanlialp2015,Lines1997,Kutnjak2015,Uchino2000,Prosandeev2008,Lisenkov2009,Ponomareva2012,Rose2012,Marathe2016,Geng2015,Defay2013,Defay2016,Defay_2016,Guzman-Verri2016,Marathe2014}
and references therein) and have led to a better knowledge of electrocaloric
effects in typical ferroelectrics, such as BaTiO$_{3}$, LiNbO$_{3}$,
Pb(Zr$_{0.4}$Ti$_{0.6}$)O$_{3}$, (Ba$_{0.5}$Sr$_{0.5}$)TiO$_{3}$,
as well as antiferroelectrics such as La-doped Pb(Zr,Ti)O$_{3}$.
On the other hand, fewer investigations about EC effects \cite{Zhang2010,Pirc2011,Pirc2011-1}
have been performed in another class of ferroelectrics, namely the
relaxor ferroelectrics. These intriguing materials exhibit unusual
features, such as a frequency-dependent and broad dielectric response
\textit{versus} temperature while remaining macroscopically paraelectric
down to 0 K \cite{Cross1994}. They also display several characteristic
temperatures (i.e., the $T_{b}$ Burns temperature, the $T^{*}$ temperature
and the $T_{m}$ temperature) that are associated with a subtle change
in some physical properties \cite{Burns1983,Dkhil2009,Svitelskiy2005,Vogel1921,Fulcher1925,Jeong2005}.
For instance, in Ba(Zr$_{0.5}$Ti$_{0.5}$)O$_{3}$ (BZT) relaxor
ferroelectrics, simulations \cite{Akbarzadeh2012} indicate that the
Burns temperature (below which the dielectric response does not obey
the Curie-Weiss law \cite{Kittel2004}) is $T_{b}\simeq450$ K , $T^{*}$
$\simeq240$ K, and $T_{m}\simeq130$ K is the temperature at which
the dielectric response exhibits a peak, as also in-line with measurements
in BZT compounds \cite{Dkhil2009,Svitelskiy2005,Maiti2008,Fahri1999}.
The microscopic origin of these features is commonly believed to be
the existence of the so-called polar nanoregions (PNRs) below the
Burns temperature \cite{Bokov2006}. Interestingly, studies devoted
to EC effects in relaxor ferroelectrics have resulted in original
findings. One example includes the failure of indirect methods (which
are based on thermodynamic equilibrium considerations) in the relaxor
ferroelectric PVDF-TrFE-CFE terpolymer to obtain the real change in
temperature induced by an electric field for temperatures below which
the broad dielectric constant peaks, because of non-ergodicity \cite{Zhang2010}.
Another example is the non-monotonic behavior of the EC coefficient
with the magnitude of the electric field at the fixed critical point
temperature $T_{CP}$ in Pb(Mg,Nb)O$_{3}$ (PMN), (Pb,La)(Zr,Ti)O$_{3}$
and Pb(Mg,Nb)O$_{3}$\textendash PbTiO$_{3}$ relaxors \cite{Pirc2011-1};
especially intriguing is the existence of a maximum of this coefficient
at the specific field ${\cal E_{CP}}$ for this $T_{CP}$ temperature,
with ($T_{CP}$, ${\cal E_{CP}}$) corresponding to the critical point
at which the paraelectric-to-ferroelectric transition changes its
nature from first order to second order. It is worthwhile to realize
that these latter results were obtained for \textit{lead-based} relaxor
ferroelectrics while there are also (environmentally-friendly) lead-free
relaxor ferroelectrics, such as Ba(Zr$_{1-x}$Ti$_{x}$)O$_{3}$,
that are fundamentally distinct. For instance, the difference in polarizability
between Ti and Zr ions in Ba(Zr$_{0.5}$Ti$_{0.5}$)O$_{3}$ was found
to be essential to reproduce relaxor behavior via the formation of
small Ti-rich PNRs embedded in a paraelectric matrix \cite{Akbarzadeh2012},
while the relaxor nature of lead-based PMN was predicted to rather
originate from a complex interplay between random electric fields,
ferroelectric and antiferroelectric interactions \textendash{} yielding
much larger PNRs touching each other at low temperatures \cite{Prosandeev2015}.
Another striking difference between Ba(Zr$_{0.5}$Ti$_{0.5}$)O$_{3}$
and PMN is that a recent atomistic simulation did not find any trace
of a \textit{first-order} paraelectric-to-ferroelectric phase transition
when subjecting Ba(Zr$_{0.5}$Ti$_{0.5}$)O$_{3}$ to electric fields,
that is, the polarization seems to always continuously evolve with
the magnitude of the $dc$ electric field in this lead-free compound
\cite{Prosandeev2013}.

One may therefore wonder about EC effects in lead-free relaxor ferroelectrics,
even more when realizing that a recent study done in Ba(Zr$_{1-x}$Ti$_{x}$)O$_{3}$
with $x=0.20$ reported a giant $\alpha$ electrocaloric coefficient
\cite{Qian2014,Ye2014} (note that this system is different from Ba(Zr$_{0.5}$Ti$_{0.5}$)O$_{3}$
in the sense that it possesses a polar ground state in addition to
some relaxor features). For instance, many questions remain to be
addressed in Ba(Zr$_{0.5}$Ti$_{0.5}$)O$_{3}$: Do indirect and direct
methods also provide different results below a specific temperature?
How does $\alpha$ behave with the $dc$ electric field for the different
temperature ranges in BZT, i.e. above $T_{b}$, between $T_{b}$ and
$T^{*}$, between $T^{*}$ and $T_{m}$, and below $T_{m}$? In particular,
can $\alpha$ exhibit a maximum for some intermediate field at any
of these temperature ranges? If such maximum exists, what is its microscopic
origin? Other natural questions to ask are if and how $\alpha$ depends
on temperature for fixed electric fields, and if it is possible to
reproduce and understand such (presently unknown) dependency.

As we will see below, this manuscript provides an answer to all these
open questions, by conducting and analyzing atomistic simulations
on Ba(Zr$_{0.5}$Ti$_{0.5}$)O$_{3}$ ferroelectric relaxors. This
article is organized as follows. Section II provides details about
the methods used here. Results are given, analyzed and explained in
Section III. Finally, Section IV concludes this work.

\section{Methods}

We use here a first-principles-based effective Hamiltonian (H$_{eff}$)
approach that has been recently developed for Ba(Zr$_{0.5}$Ti$_{0.5}$)O$_{3}$
(BZT) solid solutions \cite{Akbarzadeh2012,Prosandeev2013,Prosandeev_2013,Wang2014,Wang2016}.
The total energy of the effective Hamiltonian used here contains two
main terms: $E_{int}(\{\mathrm{\mathbf{u}}_{i}\},\thinspace\{\mathbf{v}_{i}\},\thinspace\eta_{H},\thinspace\{\sigma_{j}\})=E_{\mathrm{ave}}(\{\mathrm{\mathbf{u}}_{i}\},\thinspace\{\mathbf{v}_{i}\},\thinspace\eta_{H})+E_{\mathrm{loc}}(\{\mathrm{\mathbf{u}}_{i}\},\thinspace\{\mathbf{v}_{i}\},\thinspace\{\sigma_{j}\})$,
where $\{\mathrm{\mathbf{u}}_{i}\}$ is the local soft mode in unit
cell $i$ (which is related to the electric dipole of that cell and
that is technically centered on the Zr or Ti ions), $\{\mathbf{v}_{i}\}$
are variables related to the inhomogeneous strain inside each cell,
$\eta_{H}$ is the homogeneous strain tensor, and $\{\sigma_{j}\}$
represents the atomic configuration of the BZT solid solutions (i.e.,
how Zr and Ti ions are distributed within the B-sublattice of BZT).
$E_{\mathrm{ave}}$ contains five energetic terms: (i) the local-mode
self-energy; (ii) the long-range dipole-dipole interaction; (iii)
the energy due to short-range interactions between local modes; (iv)
the elastic energy; and (v) the energy representing the interaction
between local modes and strains \cite{Zhong1995}. $E_{loc}$ describes
how the actual distribution of Zr and Ti cations affects the energetics
involving the local soft-modes $\mathbf{u}_{i}$ and the local strain
variables, and therefore depends on the $\{\sigma_{j}\}$ distribution
\cite{Akbarzadeh2012,Prosandeev2013,Prosandeev_2013}. One can also
add to $E_{int}$ an energy given by the dot product between polarization
and electric field, in order to mimic the effect of such field on
physical properties.

This effective Hamiltonian successfully predicted the existence of
three characteristic temperatures in BZT, namely the Burns temperature
($T_{b}\simeq450$ K) below which the dielectric response does not
follow anymore the Curie-Weiss law \cite{Kittel2004}, the so-called
$T^{*}$ (that is close to $\simeq240$ K), and the $T_{m}$ temperature
at which the dielectric response can exhibit a peak ($T_{m}\simeq130$
K) \cite{Akbarzadeh2012}, as consistent with experimental findings
for BZT systems \cite{Maiti2008,Dkhil2009,Svitelskiy2005,Fahri1999}.
This atomistic scheme also yields polar nanoregions inside which the
Ti-centered dipoles are aligned parallel to each other, with these
PNRs being dynamic in nature between $T^{*}$ and $T_{b}$ while,
below $T_{m}$, they are static and all have a polarization pointing
along one of the eight $\left\langle 111\right\rangle $ pseudo-cubic
directions \cite{Akbarzadeh2012}. The polarizations of these different
PNRs cancel each other, as consistent with the fact that BZT is macroscopically
paraelectric down to 0 K \cite{Akbarzadeh2012,Maiti2008,Dkhil2009,Svitelskiy2005,Fahri1999}.
This effective Hamiltonian was also successful in reproducing the
unusual dielectric relaxation known to occur in relaxor ferroelectrics
\cite{Wang2016}. Here, we implement this H$_{eff}$ within Monte
Carlo (MC) and Molecular Dynamics (MD) simulations, in order to determine
and understand EC effects in BZT relaxors \textendash{} as modeled
by $14\times14\times14$ supercells (13720 atoms) in the MC computations
and $32\times32\times32$ (32768 atoms) in the MD simulations. Note
that this different choice of supercells between the MC and MD simulations
originates from the fact that the code we used for the MD computations
can handle larger supercells, and that the use of $32\times32\times32$
supercells allows the temperature change in MD simulations to be easily
sorted out from the temperature fluctuations. Note also that we numerically
checked that the use of $12\times12\times12$, $14\times14\times14$
and $16\times16\times16$ supercells provides similar results, which
suggests that our Monte-Carlo simulations are free from significant
size effects. These supercells are periodic along the three Cartesian
directions, and Zr and Ti atoms are randomly distributed inside them.
We also average our physical results over 20 of these random configurations
for both MC and MD simulations, in order to mimic well disordered
BZT solid solutions.

Let us now indicate how we practically compute, from these simulations,
the electrocaloric coefficient $\alpha=\left.\frac{\partial T}{\partial{\cal E}}\right|_{S}$.
One approach we use here is based on the Maxwell thermodynamical relationship
$\left.\frac{\partial S}{\partial{\cal E}}\right|_{T}=\left.\frac{\partial P}{\partial T}\right|_{{\cal E}}$
leading to the adiabatic temperature change 
\begin{equation}
\Delta T=-\intop_{{\cal E}_{1}}^{{\cal E}_{2}}\frac{T({\cal E)}}{C_{{\cal E}}(T)}\left.\frac{\partial P}{\partial T}\right|_{{\cal E}}d{\cal E},
\end{equation}
where $P$ is the macroscopic polarization and $C_{{\cal E}}$ is
the heat capacity per unit volume under constant $dc$ electric field.
Such latter equation therefore tells us that we can obtain $\alpha$
from MC simulations by computing

\begin{equation}
\alpha=-\frac{T}{C_{{\cal E}}}\left.\frac{\partial P}{\partial T}\right|_{{\cal E}}.\label{eq:ece-alpha}
\end{equation}
This way of extracting $\alpha$ is coined MC-1 here.

\begin{figure}
\includegraphics[width=8cm]{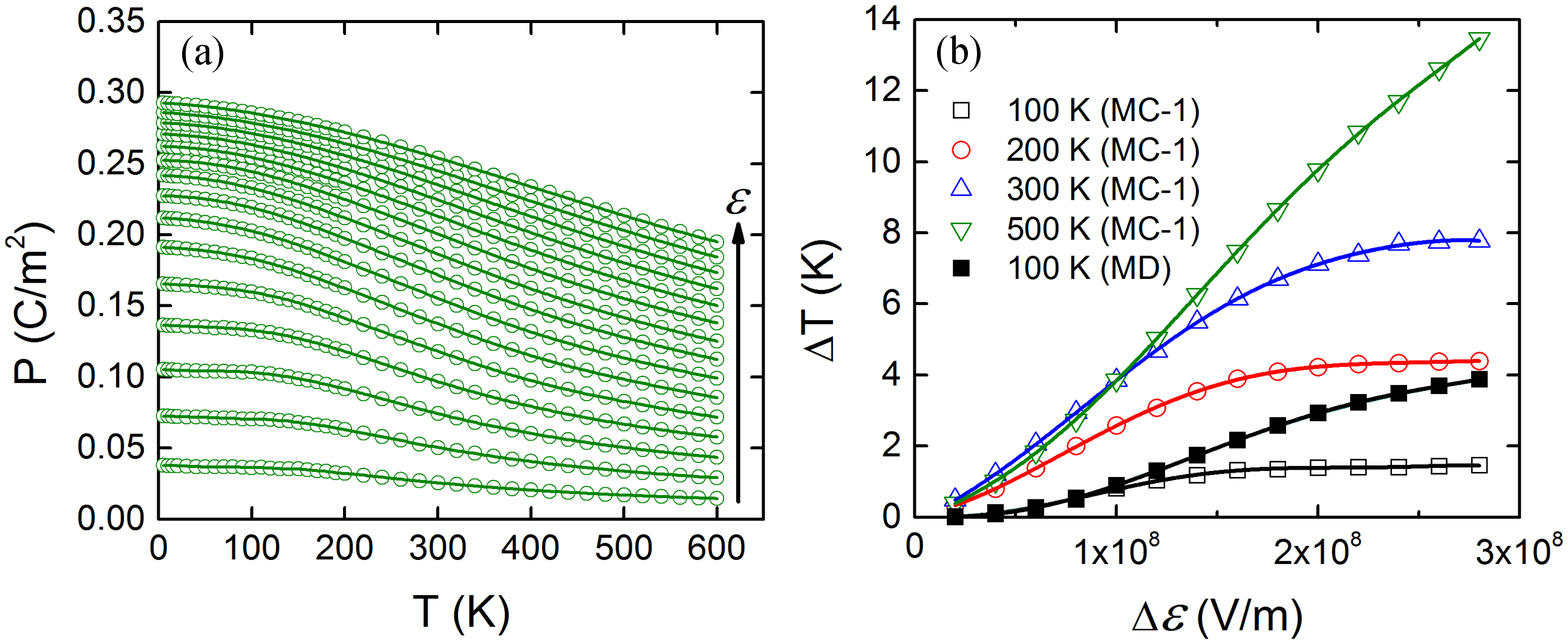}

\caption{(Color online) Physical properties associated with the MC-1 method.
Panel (a) shows the temperature dependency of the polarization in
BZT systems subject to different $dc$ electric fields, all applied
along the pseudo-cubic {[}001{]} direction but varying from $2.0\times10^{7}$
to $3.0\times10^{8}$ V/m in magnitude by steps of $2.0\times10^{7}$
V/m. Panel (b) shows the resulting change in temperature as a function
of $\Delta{\cal E}$=${\cal E}_{2}$-${\cal E}_{1}$ for four selected
initial temperatures, as computed from Eq. (1) and choosing ${\cal E}_{1}=2.0\times10^{7}$
V/m. Note that Panel (b) also further reports the direct change in
temperature at 100 K as a function of ${\cal E}_{f}$. \label{fig:deltaT_vs_T_MC}}
\end{figure}

For instance, Fig. \ref{fig:deltaT_vs_T_MC}(a) reports the polarization
as a function of temperature obtained from MC simulations on Ba(Zr$_{0.5}$Ti$_{0.5}$)O$_{3}$,
for $dc$ electric fields all applied along the pseudo-cubic {[}001{]}
direction and ranging between $2.0\times10^{7}$ and $3.0\times10^{8}$
V/m in magnitude. Values of $\left.\frac{\partial P}{\partial T}\right|_{{\cal E}}$
are then obtained from cubic B-spline fits to these $P(T)$ curves,
which allows us to determine $\alpha$ via Eq. (\ref{eq:ece-alpha}).
Note that the heat capacity at a given electric field ${\cal E}$
is calculated as: $C_{{\cal E}}=(N\frac{\left\langle {E_{int}}^{2}\right\rangle -\left\langle {E_{int}}\right\rangle ^{2}}{T^{2}k_{B}}+\frac{15}{2}k_{B})/V$,
where $N$ is the number of sites in the supercell, ${E_{int}}$ is
the total internal energy provided by the effective Hamiltonian, $\left\langle \ \right\rangle $
denotes the average over the MC sweeps at every considered $T$ temperature,
$k_{B}$ is the Boltzmann constant, and $V$ is the volume of the
unit cell. The factor $\frac{15}{2}$ in that formula reflects that
there are five atoms in the unit cell of perovskites \cite{Ponomareva2012}.
Moreover, $C_{{\cal E}}$ is computed for different temperatures and
electric fields, implying that it can, in principle, depend on $T$
and ${\cal E}$. However, we numerically found that these dependencies
are rather weak as consistent with measurements \cite{Qian2014} and
that $C_{{\cal E}}$ is always very close to 2.18 MJ/K m$^{3}$.

Interestingly, there is another way to obtain the EC coefficient from
MC runs, that is by taking advantage of the cumulant formula given
in Ref. \cite{Omran2016}: 
\begin{equation}
\alpha=-\thinspace Z^{*}a_{lat}NT\thinspace\{\frac{\left\langle \mathbf{\left|u\right|}{E_{int}}\right\rangle -\left\langle \mathbf{\left|u\right|}\right\rangle \left\langle {E_{int}}\right\rangle }{\left\langle {E_{int}}^{2}\right\rangle -\left\langle {E_{int}}\right\rangle ^{2}}\},\label{eq:ece-alphaMC2}
\end{equation}
where $Z^{*}$ is the Born effective charge, $a_{lat}$ is the five-atom
lattice constant, $N$ is the number of sites in the supercell, $T$
is the considered temperature, $\mathbf{u}$ is the supercell average
of the local mode, ${E_{int}}$ is the total energy of the effective
Hamiltonian, and $\left\langle \ \right\rangle $ denotes the average
over the MC sweeps at every considered temperature. This method will
be called MC-2 here. Technically, the computation of $\alpha$ via
Eq. (\ref{eq:ece-alphaMC2}) is done for a chosen combination of temperature
and magnitude of a $dc$ electric field applied along the pseudo-cubic
{[}001{]} direction, which therefore allows us to determine the effect
of temperature and applied electric field on the EC coefficient. In
the following, we will also be interested in comparing the predictions
of MC-1 and MC-2, mostly because the MC-2 method is less known than
MC-1 while being computationally more accurate (since, unlike MC-1,
it does not rely on a fit of $\left.\frac{\partial P}{\partial T}\right|_{{\cal E}}$).

Regarding the direct approach, we determine the electrocaloric coefficient
by using the ramping method of Ref. \cite{Marathe2016} within Molecular
Dynamics. First, an Evans-Hoover thermostat \cite{Hoover1982,Evans1983}
is used in the MD simulations in order to equilibrate the system at
an initial temperature $T$ when no electric field is applied. The
electric field is then applied along the pseudo-cubic {[}001{]} direction
and ramped up (with time) from zero to a specific value, ${\cal E}_{f}$,
and then ramped down from ${\cal E}_{f}$ to zero. Practically, we
chose the time dependence of the applied field $\mathcal{E}(t)$ amplitude
to be 

\begin{equation}
\mathcal{E}(t)=\frac{\mathcal{E}_{f}}{2}\left(\tanh\left(\frac{t-t_{\text{up}}}{\tau}\right)-\tanh\left(\frac{t-t_{\text{down}}}{\tau}\right)\right),
\end{equation}
where $t_{\text{up}}$ and $t_{\text{down}}$ denote the times when
the field magnitude reaches $\mathcal{E}_{f}/2$ during ramping up
and down, respectively. The ramping up/down time frames thus correspond
to

\begin{equation}
t_{\text{up/down}}~-~\tau/2\lesssim~t~\lesssim~t_{\text{up/down}}~+~\tau/2,
\end{equation}
with $\tau$ representing the time interval during which the field
on/off switching happens. The ``hyperbolic tangent'' time profile
is commonly used in linear response calculations and was chosen to
obtain a smooth time dependence of the external field. Notably we
observed no significant differences with test calculations where the
time dependence of the external field was assumed linear as described
in Ref. \cite{Marathe2016}. To test the convergence of results with
respect to $\tau$, and the integration time-step $\Delta t$, the
test runs were performed for values of $\tau$ ranging from 20 ps
to 200 ps and values of $\Delta t$ from 0.001 fs to 4 fs. All the
simulation were performed using the Omelyan second order symplectic
integration algorithm \cite{Omelyan2003}. Based on the convergence
tests, the final chosen value of $\tau$ was of 188 ps with $\Delta t$
equal to 0.1 fs ensuring the energy conservation for constant field
simulation up to the maximum relative error of 10$^{-6}$. The inverse
rate of the change of the applied field was thus close to 188 $\text{fs}\cdot\text{cm}/\text{kV}$
for the applied field magnitude of 1000 $\text{kV}/\text{cm}$. For
the chosen simulation parameters, we find that the calculated field
induced temperature change upon ramping down $\Delta T_{\text{down}}$
is equal in magnitude, but opposite in sign, to the temperature change
$\Delta T_{\text{up}}$ produced by the switching on the external
field for temperatures above $T_{m}$ \textemdash{} a result that
is naturally expected for time-reversible processes. However, for
$T<T_{m}$, during the ramping down of the applied field the temperature
first exhibited a drop which was subsequently followed by an increase
(note that this result was also tested for convergence with respect
to $\tau$ and $\Delta t$). Such behavior, broadly speaking, can
be attributed to the loss of ergodicity below $T_{m}$. The detailed
investigation of the microscopic mechanism responsible for this unusual
behavior lies beyond the scope of the current study and, for the purposes
of the present work, the EC temperature change $\Delta T$ was defined
to be equal to $\Delta T_{\text{up}}$, and the $\alpha$ EC coefficient
associated with a specific field's magnitude can then be obtained
by taking the derivative of $\Delta T_{\mathrm{up}}$ with respect
to ${\cal E}_{f}$ at this specific field's magnitude. Such results
will be denoted as ``MD'' here \cite{footnoterenrmalizationMD}.

Note that data from MC-1 and MC-2 approaches can be considered to
be associated with the \textit{indirect method} to obtain EC effects,
because they are based on thermodynamic equilibrium. On the other
hand, data obtained from MD computations yield the \textit{direct}
EC effects, which may differ from those obtained from the indirect
way for systems adopting non-ergodic behavior, as the one that relaxors
are known to exhibit below some specific temperature $T_{m}$ at which
the dielectric response peaks \cite{Lu2010}. Comparisons between
our MC and MD results should thus tell us the difference between the
indirect and direct ways to extract EC effects in relaxors. Since
we are also interested in checking if and how this difference (if
any) depends on the investigated temperature region, we decided to
focus on four particular representative temperatures. They are: (1)
500 K, which is above the predicted Burns temperature ($T_{b}\simeq450$
K) of BZT \cite{Akbarzadeh2012,Maiti2008}; (2) 300 K, which is located
in-between our critical $T^{*}\simeq240$ K \cite{Akbarzadeh2012,Dkhil2009,Svitelskiy2005}
and $T_{b}$; (3) 200 K, that is now between the computed $T_{m}$
temperature of BZT ($T_{m}\simeq130$ K) \cite{Akbarzadeh2012,Fahri1999}
and $T^{*}$; and (4) 100 K, which is thus below $T_{m}$ (note that
the Supplemental Material \cite{Supplemental Material} also shows
our results for the EC coefficient in BZT at 600 K).

\section{Results }

\subsection{EC coefficients}

\begin{figure}
\includegraphics[width=8cm]{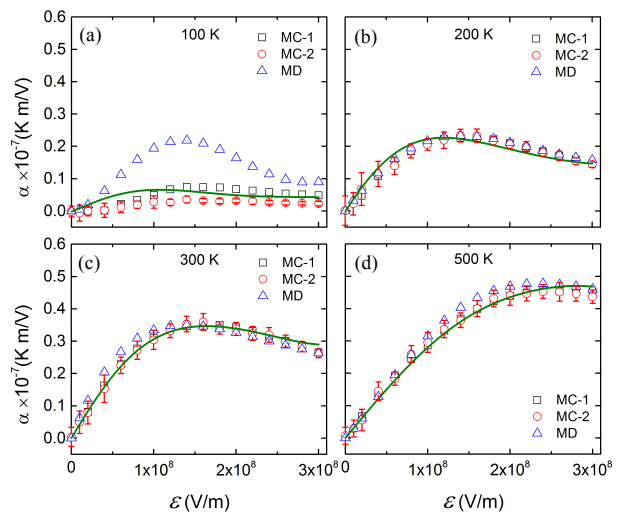}

\caption{(Color online) Electrocaloric coefficient, $\alpha$, as a function
of the applied $dc$ electric field ${\cal E}$, as predicted for
the different indirect and direct approaches at 100 K, 200 K, 300
K and 500 K (Panels (a)-(d), respectively). The solid green line represent
the fit of the MC-1 and MC-2 results by the second line of Eq. (9),
i.e., $\alpha=\beta T\left.\frac{\partial P^{2}}{\partial{\cal E}}\right|_{T}$,
where $\beta$ is a constant and $\left.\frac{\partial P^{2}}{\partial{\cal E}}\right|_{T}$
is obtained from the data of Fig. \ref{fig:P_square_vs_E}. Error
bars (resulting from the use of 20 different disordered alloy configurations)
are also shown for the MC-2 data. \label{fig:alpha_vs_E}}
\end{figure}

Figure \ref{fig:alpha_vs_E} shows the electrocaloric coefficient
as a function of electric field, ${\cal E}$, for these four different
selected temperatures, and as computed from the aforementioned MC-1,
MC-2 and MD methods. One can first clearly see that, for any of these
temperatures, the (indirect) MC-1 and MC-2 approaches provide nearly
identical results. Similarly, $\alpha$ predicted by the (direct)
MD scheme agrees very well with those of MC-1 and MC-2 for 200 K,
300 K and 500 K at any field, which demonstrates that indirect methods
based on Maxwell thermodynamic relation can be safely used to estimate
$\alpha$ above the $T_{m}$ temperature of relaxors. On the other
hand, Fig. \ref{fig:alpha_vs_E}(a) clearly reveals that the EC coefficient
of the MD method significantly differs from that predicted by MC-1
and MC-2 at 100 K, as a result of non-ergodicity. In particular, at
100 K, the $\alpha$ deduced from the indirect methods are smaller
than that those directly extracted, which is in agreement with previous
reports \cite{Zhang2010,Goupil2012,Lu2010}. It is also interesting
to realize that the EC coefficient of the MD method gets closer to
those of MC-1 and MC-2 at 100 K for the highest considered electric
fields. This is because, under high electric fields, BZT relaxors
can be converted to a normal ferroelectric and thus becomes ergodic
\cite{Prosandeev2013}.

Moreover, the results of Fig. \ref{fig:alpha_vs_E}(d) also indicate
that $\alpha$ at 500 K is vanishing at small fields and then increases
with ${\cal E}$, until it very slightly decreases for our highest
investigated fields. Interestingly, our values of $\alpha$ for high
fields at 500 K are of the order of $0.5\times10^{-7}$ K m/V, that
is similar to the predicted one of $0.67\times10^{-7}$ K m/V in a
ferroelectric phase of (Ba,Sr)TiO$_{3}$ \cite{Lisenkov2009}. Figures
\ref{fig:alpha_vs_E}(a), \ref{fig:alpha_vs_E}(b) and \ref{fig:alpha_vs_E}(c)
also show that, for temperatures below the Burns temperature, $\alpha$
adopts a very clear \textit{maximum} for an intermediate field (whose
value is dependent on temperature) within our investigated range of
electric fields. In other words, at temperatures of 300 K, 200 K or
100 K, the EC coefficient first increases with field before noticeably
decreasing. Such non-mononotic behavior of $\alpha$ (starting with
a vanishing value at small fields and having a peak for an intermediate
field before decreasing for larger fields) was indeed measured, as
well as reproduced by the so-called phenomenological spherical random
bond random field model, in Pb(Mg,Nb)O$_{3}$, (Pb,La)(Zr,Ti)O$_{3}$
and Pb(Mg,Nb)O$_{3}$\textendash PbTiO$_{3}$ relaxors in Ref. \cite{Pirc2011-1},
but only for a specific temperature: namely, the critical temperature
at which the discontinuous electric-field-induced ferroelectric transition
of these systems becomes continuous (for the value of the electric
field associated with the maximum of $\alpha$). Our results displayed
in Fig. \ref{fig:alpha_vs_E} therefore generalize such finding by
indicating that, for \textit{any temperature}, $\alpha$ of BZT can
also exhibit a maximum within the investigated field range. Further,
note also that BZT differs from the cases of Pb(Mg,Nb)O$_{3}$, (Pb,La)(Zr,Ti)O$_{3}$
and Pb(Mg,Nb)O$_{3}$\textendash PbTiO$_{3}$ in the sense that the
temperature behavior of the polarization displayed in Fig. \ref{fig:deltaT_vs_T_MC}(a)
is always continuous for any investigated field. It is worthwhile
to know that the maximum of $\alpha$ at a certain field was also
predicted to occur in Ba$_{0.5}$Sr$_{0.5}$TiO$_{3}$ \cite{Ponomareva2012}
and defect doped BaTiO$_{3}$ \cite{Ma2016}, and that we also found
this non-mononotic behavior of $\alpha$ in the paraelectric phase
of BaTiO$_{3}$ (BTO) bulk \textendash{} as evidenced in the Supplemental
Material \cite{Supplemental Material}.

\subsection{Analysis of the results via a Landau-like model}

Let us now try to understand the main results of Fig. \ref{fig:alpha_vs_E}.
For that, we start from a simplest Landau free-energy potential describing
the behavior of a non-linear dielectric 
\begin{equation}
\begin{split}F & =F_{0}(T)+\Delta F(T,P,{\cal E})\\
 & =F_{0}(T)+\frac{1}{2}a(T)P^{2}+\frac{1}{4}bP^{4}-{\cal E}P,
\end{split}
\label{eq:Landau expansion}
\end{equation}
where $F_{0}(T)$ captures the basic temperature dependence of the
free energy of the materials, and the other terms account for the
variations that involve the development of a polarization or application
of an electric field. Note that the temperature dependence of the
harmonic $a(T)$ parameter can be a complex one in our BZT compound
with various regimes, as inferred from the temperature behavior of
the dielectric response under \textit{dc} field and discussed in Ref.
\cite{Akbarzadeh2012}: for $T>T_{b}$ we have $a(T)\propto(T-T_{0})$,
while for $T<T_{m}$ we have $da(T)/dT\sim0$, and for $T_{m}<T<T_{b}$
we have a smooth interpolation between these two regimes (note that
(i) $T_{0}$ is extracted from the Curie-Weiss behavior of the dielectric
response above $T_{b}$ and can be negative in relaxor ferroelectrics,
as predicted and experimentally found in Refs. \cite{Akbarzadeh2012,Maiti2008}
; and (ii) that the aforementioned behaviors of $a(T)$ implies that
it is increasing with temperature above $T_{m}$). In the following
equations we will work with a generic $a(T)>0$, noting that the final
results have to be interpreted depending on the $T$ region we are
in. In particular, the phenomenological equations to be derived here
(namely, Eqs. (6)-(16)) can only be safely applied to temperatures
above $T_{m}$. This is because these equations rely on thermodynamic
equilibrium while BZT is non-ergodic below $T_{m}$. Finally, the
positive parameter $b>0$ accounts for the saturation of the dielectric
response of the material.

Let us now discuss the behavior of the EC coefficient as predicted
by this simple model. The entropy can be obtained as 
\begin{equation}
S=-\frac{dF}{dT}=-\frac{dF_{0}}{dT}-\frac{\partial\Delta F}{\partial T}-\frac{\partial\Delta F}{\partial P}\frac{dP}{dT}.
\end{equation}
Noting that at equilibrium we have $\partial\Delta F/\partial P=0$,
we obtain: 
\begin{equation}
S=-\frac{dF_{0}}{dT}-\frac{a'(T)}{2}P^{2},
\end{equation}
where $a'=da/dT$. It is then straightforward to derive the following
expression for $\alpha$: 
\begin{equation}
\begin{split}\alpha & =-\left.\frac{T}{C_{{\cal E}}}\frac{\partial S}{\partial{\cal E}}\right|_{T}\\
 & =\left.\frac{Ta'(T)}{2C_{{\cal E}}}\frac{\partial P^{2}}{\partial{\cal E}}\right|_{T}\\
 & =\frac{Ta'(T)}{C_{{\cal E}}}P\chi,
\end{split}
\end{equation}
where $\chi$ is the dielectric susceptibility.

Interestingly, the behavior of a dielectric for small electric fields
can be readily discussed from this expression. Indeed, if $P=0$ for
${\cal E}=0$, then we have $P=\chi{\cal E}$, which leads to $\alpha\propto{\cal E}$,
assuming that the dependence of the specific heat $C_{{\cal E}}$
on the electric field can be neglected. This prediction is fully consistent
with the null value of $\alpha$ reported in Fig. \ref{fig:alpha_vs_E}
at zero field for any temperature, and immediately implies that $\Delta T\propto{\cal E}^{2}$
\textendash{} which shows that the EC effect is null in the limit
of small ${\cal E}$.

To discuss the behavior of $\alpha$ for arbitrary electric-field
values, we recall the equilibrium condition $\partial F/\partial P=0$
to obtain 
\begin{equation}
a(T)P+bP^{3}={\cal E}.
\end{equation}
Further, if we take the derivative with respect to the electric field
on both sides of this equation, we get 
\begin{equation}
a(T)\chi+3bP^{2}\chi=1,\label{eq:chi}
\end{equation}
which leads to 
\begin{equation}
\alpha=\frac{2T}{C_{{\cal E}}}\frac{a'(T)P}{a(T)+3bP^{2}}.
\end{equation}
This interesting expression implies that, in the limit of large polarizations
(or, equivalently, large electric fields), we have $\alpha\rightarrow0$.
Hence, since we also know that $\alpha=0$ for ${\cal E}=P=0$, it
immediately follows that the EC coefficient will present at least
one extremum (maximum or minimum) at intermediate values of the electric
field, as also consistent with our numerical results of Fig. \ref{fig:alpha_vs_E}.
Of course, whether or not such an extremum is experimentally accessible
will depend on the breakdown field of a particular material or sample;
yet, at least one extremum has to exist in principle. Note also that
$\alpha$ will adopt a maximum if $a'(T)$ is positive (which is the
case of BZT) while it will possess a minimum if $a'(T)$ is negative.

To find the electric field that makes $\alpha$ maximum, we have to
solve 
\begin{equation}
\frac{d\alpha}{d{\cal E}}=-\frac{2a'(T)}{C_{{\cal E}}}(\chi^{2}+P_{{\rm m}}\chi')=0,
\end{equation}
where $\chi'=d\chi/d{\cal E}$ captures the non-linear dielectric
response of the material, and $P_{{\rm m}}$ is the value of the polarization
for which $\alpha$ is maximum. The non-linear response $\chi'$ is
related to $P$ and $\chi$ by 
\begin{equation}
a(T)\chi'+6bP\chi^{2}+3bP^{2}\chi'=0,
\end{equation}
which we obtain by taking the field derivative of both sides of Eq.~(\ref{eq:chi}).
From the last two relations, one can show that the condition to have
an extremum of $\alpha$ reduces to 
\begin{equation}
P_{{\rm m}}^{2}=\frac{a(T)}{3b},
\end{equation}
from which several conclusions can be immediately drawn. First, for
stiff materials \textendash{} i.e., those with $a(T)\gg0$ \textendash{}
the extremum of $\alpha$ will occur at relatively large value of
the polarization and applied electric field. Similarly, if the dielectric
response is very linear \textendash{} i.e., for small $b>0$ \textendash ,
the extremum of $\alpha$ will also tend to occur for large values
of $P$ and ${\cal E}$. Finally, using a linear approximation for
the polarization as a function of field, $P\sim\chi{\cal E}$, we
can write 
\begin{equation}
{\cal E}_{{\rm m}}^{2}\approx\frac{a(T)}{3b\chi^{2}}=\frac{4a^{3}(T)}{3b},
\end{equation}
which provides us with a useful (albeit approximate) expression for
the electric field corresponding to $\alpha$'s extremum. For instance,
it tells us that ${\cal E}_{{\rm m}}$ should increase with temperature
if $a(T)$ is enhanced with temperature (which is precisely the case
for BZT). This increase of ${\cal E}_{{\rm m}}$ with temperature
is indeed confirmed in Fig. \ref{fig:alpha_vs_E} for temperatures
above 200 K, and is also consistent with the fact that, at 500 K,
the maximum of $\alpha$ occurs for electric fields being close to
our highest investigated values.

Moreover, the second line of Eq. (9) indicates that $\alpha=\beta T\left.\frac{\partial P^{2}}{\partial{\cal E}}\right|_{T}$,
with $\beta=\frac{a'(T)}{2C_{{\cal E}}}$. In other words, assuming
that $C_{{\cal E}}$ is independent of temperature and electric field,
and that $a'(T)$ is also a constant (which is, e.g., what Curie-Weiss
law \cite{Kittel2004} provides), this expression implies that the
numerical data of the MC-1 and MC-2 approaches for the EC coefficient
should be well fitted by the product of temperature and the derivative
of the square of the polarization with respect to electric field,
once rescaling this product by a constant \cite{footnoteequation7,Strukov1967}.
Figure \ref{fig:alpha_vs_E} indeed tells us that this is the case
for any temperature (especially at and above 200 K, where we are in
ergodic equilibrium conditions), since these figures further display
the results of such fits by means of solid green curves. In other
words, one can safely use Eq. (9) to reproduce and understand the
EC coefficients numerically obtained by the indirect methods for any
temperature and field (note that the Supplemental Material also shows
that Eq. (9) can be accurately used for the $\alpha$ coefficient
of typical ferroelectrics, such as BaTiO$_{3}$, which further emphasizes
its generality). In particular, the second line of Eq. (9) indicates
that, for a given temperature, the non-monotonic and unusual behavior
of $\alpha$ with fields obtained by MC-1 and MC-2 should be directly
related to the dependence of $\left.\frac{\partial P^{2}}{\partial{\cal E}}\right|_{T}$
with ${\cal E}$. To check such interesting idea, Figs. \ref{fig:P_square_vs_E}(a)-\ref{fig:P_square_vs_E}(d)
report the square of the macroscopic polarization as a function of
electric field applied along the {[}001{]} direction at 100 K, 200
K, 300 K and 500 K, respectively. The central inset of these figures
displays the derivative of this quantity with respect to the field,
and reveal that, indeed, $\left.\frac{\partial P^{2}}{\partial{\cal E}}\right|_{T}$
has the same trend as the indirect EC coefficient of Fig. \ref{fig:alpha_vs_E}.
In particular, Figs. \ref{fig:P_square_vs_E}(a)-\ref{fig:P_square_vs_E}(d)
reveal that $\alpha$ is very small for low fields at any temperature,
simply because the square of the polarization is basically independent
of electric fields for small ${\cal E}$ \cite{squareP}. Such strong
connection between $\alpha$ and $\left.\frac{\partial P^{2}}{\partial{\cal E}}\right|_{T}$
is reinforced when realizing that the field resulting in a maximum
of the $\alpha$ coefficient of the MC-1 and MC-2 methods at 100 K,
200 K, 300 K and 500 K is very close to the field at which $\left.\frac{\partial P^{2}}{\partial{\cal E}}\right|_{T}$
is optimal at these temperatures. It is also interesting to realize
that the maximal value of the $\alpha$ of the indirect methods increases
by a factor of about 3 when increasing the temperature from 100 K
to 300 K, while the corresponding maximum of $\left.\frac{\partial P^{2}}{\partial{\cal E}}\right|_{T}$
is quite similar between 100 K and 300 K. Such feature can, in fact,
be understood by the fact that the second line of Eq. (9) indicates
that the EC coefficient is directly proportional to the temperature.
In other words, increasing the temperature increases $\alpha$ in
case of similar $\left.\frac{\partial P^{2}}{\partial{\cal E}}\right|_{T}$
(note that Eq. (9) is also consistent with the computational finding
of the enhancement of $\alpha$ with temperature in the ferroelectric
phases of (Ba,Sr)TiO$_{3}$ in Ref. \cite{Lisenkov2009}).

\subsection{Microscopic insights}

\begin{figure}
\includegraphics[width=8cm]{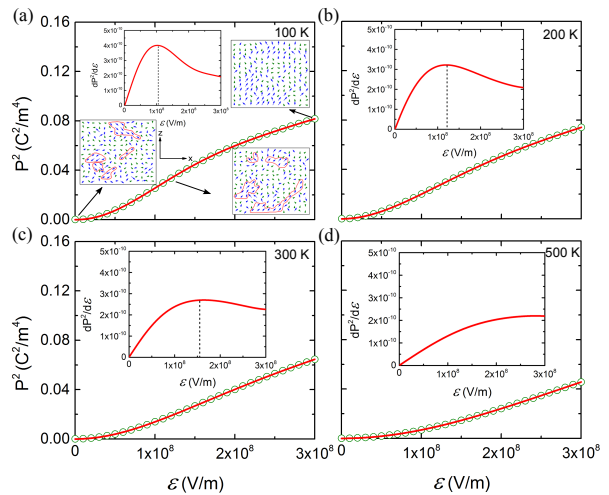}

\caption{(Color online) The square of the macroscopic polarization as a function
of the applied $dc$ electric field, at 100 K, 200 K, 300 K and 500
K (Panels (a)-(d), respectively). The red line represents a fit by
7$^{th}$ degree polynomials, which were then used to calculate the
derivative $dP^{2}/d{\cal E}$ that is shown in the corresponding
central inset of each panel. The other insets of Panel (a) show the
dipolar configurations in a given $(x,\thinspace z)$ plane at 100
K, as obtained from MC simulations for different $dc$ electric fields
(0 V/m, $1.2\times10^{8}$ V/m and $3.0\times10^{8}$ V/m) applied
along the pseudo-cubic {[}001{]} direction. In these latter insets,
the blue and green colors indicate that the local dipoles are centered
on Ti and Zr ions, respectively, and the red solid lines delimit the
PNRs. \label{fig:P_square_vs_E}}
\end{figure}

Let us now try to reveal the \textit{microscopic} origins of the maximum
of $\left.\frac{\partial P^{2}}{\partial{\cal E}}\right|_{T}$ at
200 K and 300 K (which explains the maximum of the indirect and direct
$\alpha$ of these temperatures) as well as the peak of the $\alpha$
obtained by the MD simulations at 100 K (recall that, for temperature
below $\simeq$ 130 K, BZT is non-ergodic and thus can not be technically
described by Eq. (9)). For that, we focus on the field evolution of
the microscopic configurations of BZT at 100 K. Some insets of Fig.
\ref{fig:P_square_vs_E}(a) show dipolar snapshots within a given
$(x,\thinspace z)$ plane obtained from MC simulations at 100 K for
different electric fields. They reveal that the microscopic dipolar
pattern is rather complex and sensitive to electric fields. For instance,
there are different polar nanoregions inside which the dipoles centered
on Ti ions align along one of the eight \textbf{$\left\langle 111\right\rangle $}
pseudocubic directions (with this direction varying from one PNR to
another, e.g. from {[}111{]} to {[}11$\bar{1}${]}), when no external
field is applied {[}see left bottom inset of Fig. \ref{fig:P_square_vs_E}(a){]}.
Increasing the electric field then leads to the local dipoles of the
PNRs rotating towards the field's direction, as well as the formation
of rather large PNRs having local dipoles lying along the applied
electric field direction {[}see bottom right inset of Fig. \ref{fig:P_square_vs_E}(a){
for a field of $1.2\times10^{8}$ V/m{]}}. Finally, Fig. \ref{fig:P_square_vs_E}(a)
further indicates that increasing the field up to our considered maximum
value ${\cal E}=3.0\times10^{8}$ V/m causes nearly all Ti-centered
local dipoles to align along the field's direction, which can be seen
as indicative that BZT is converting from a relaxor behavior to a
normal ferroelectric {[}see the top right inset of Fig. \ref{fig:P_square_vs_E}(a){]}.

\begin{figure}
\includegraphics[width=8cm]{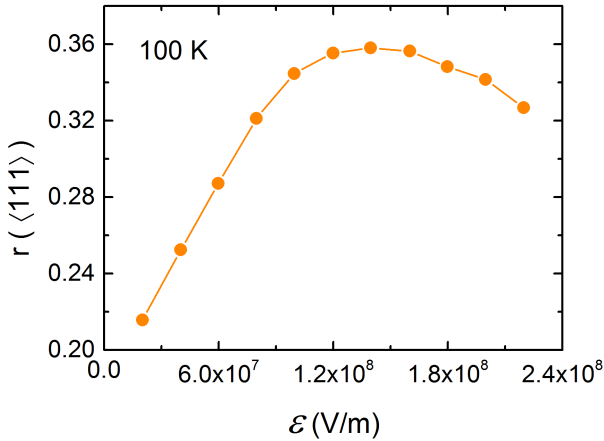}

\caption{(Color online) The ratio of dipoles that are pointing along \textbf{$\left\langle 111\right\rangle $}
directions having a positive $z$ component, as a function of the
magnitude of the electric field applied along the pseudo-cubic {[}001{]}
direction at 100 K. Note that these \textbf{$\left\langle 111\right\rangle $}
directions are thus away from the {[}001{]} field's direction. \label{fig: microscopic}}
\end{figure}

Interestingly, the aforementioned field-induced rearrangement of the
local dipoles for fields close to $1.2\times10^{8}$ V/m generates
a maximal change of the entropy, as evidenced by the fact that Fig.
\ref{fig: microscopic} reveals that the fields associated with maximal
values of $\alpha$ obtained by the direct approach at 100 K {[}see
Fig. \ref{fig:alpha_vs_E} (a){]} are precisely the fields for which
a specific microscopic feature occurs: the number of dipoles pointing
along \textbf{$\left\langle 111\right\rangle $} pseudocubic directions
for which the $z$-component is positive (i.e., which have a $z$
component parallel to the applied electric field) is maximal for these
fields. This microscopic feature was also numerically found (not shown
here) for the fields associated with the maximum values of $\alpha$
at 200 K and 300 K (note that BZT does not possess any PNR at 500
K because this latter temperature is above the Burns temperature).

\subsection{Resulting change in temperature}

Let us now concentrate on the $\Delta T$ change in temperature, associated
with the EC coefficient and as computed from Eq. (1), for the four
studied temperatures of 100 K, 200 K, 300 K and 500 K. Note that,
unlike for 200 K, 300 K and 500 K, this change in temperature will
not be the ``direct'' one for 100 K because the system is non-ergodic
at this temperature, while Eq. (1) assumes thermodynamic equilibrium.
We nevertheless report in Fig. \ref{fig:deltaT_vs_T_MC}(b) the data
for $\Delta T$ as a function of a change in electric field, $\Delta{\cal E}$,
at 100 K, along with those of 200 K, 300 K and 500 K, for the sake
of comparison. Technically, the $\Delta T$ of Eq. (1) is computed
by integrating the $\alpha$ coefficient calculated by the MC-1 indirect
method (see Eq. (2)) from ${\cal E}_{1}$ to ${\cal E}_{2}$, with
$\Delta{\cal E}$ being the difference between the magnitude of these
two fields and always choosing ${\cal E}_{1}=2.0\times10^{7}$ V/m
while varying ${\cal E}_{2}$ when changing $\Delta{\cal E}$. Two
main features can be seen from Fig. \ref{fig:deltaT_vs_T_MC}(b):
(i) for any temperature, $\Delta T$ is not linear with $\Delta{\cal E}$,
as also observed near 310 K in the Ba(Zr$_{0.2}$Ti$_{0.8}$)O$_{3}$
material \cite{Qian2014} exhibiting relaxor behavior and which is
in contrast with, e.g., the cases of the ferroelectric Pb(Zr$_{0.95}$Ti$_{0.05}$)O$_{3}$,
Pb(Zr$_{0.4}$Ti$_{0.6}$)O$_{3}$, (Ba$_{0.5}$Sr$_{0.5}$)TiO$_{3}$
and Pb(Mg,Nb)O$_{3}$-PbTiO$_{3}$ systems reported in Refs. \cite{Prosandeev2008,Lisenkov2009,Zhang2006,Saranya2009};
and (ii) for any given electric field above $\simeq$ $1.5\times10^{8}$
V/m, $\Delta T$ is enhanced when the considered initial temperature
increases. Item (i) originates from the fact that $\alpha$ strongly
depends on electric field and can even be non-mononotic with ${\cal E}$
in relaxor ferroelectrics (see Fig. \ref{fig:alpha_vs_E}). Item (ii)
can be simply understood by realizing that Eq. (9) provides a dependence
of the EC coefficient on temperature. Note that we also numerically
checked that our $\Delta T$ are not directly proportional to the
power 2/3 of the electric field, except for fields above 10$^{8}$
V/m at 500 K, which contrasts with the prediction of Ref. \cite{Guzman-Verri2016}.
Furthermore, our MD predictions for $\Delta T$ at 100 K are also
given for comparison in Fig. \ref{fig:deltaT_vs_T_MC}(b), which demonstrates,
once again, that results from direct and indirect approaches differ
below $T_{m}$. One should also recall that atomic schemes, such as
effective Hamiltonians, typically provide an overestimation by one
order of magnitude with respect to experiments for electric fields
\cite{Xu2017} while they tend to yield correct values for the EC
coefficient (as shown in the Supplemental Material). Experiments are
thus called for to determine by which factors the temperatures and
fields of Fig. \ref{fig:deltaT_vs_T_MC}(b) would have to be rescaled
in BZT (if any).

\section{Summary}

In summary, we combined an atomistic effective Hamiltonian scheme
with Monte-Carlo and Molecular Dynamics techniques to investigate
electrocaloric effects in the lead-free BZT systems subject to electric
fields of different magnitude and all oriented along the pseudo-cubic
{[}001{]} direction. It is found that, for any temperature, $\alpha$
exhibits a non-monotonic behavior with field that consists of small
values at low fields, followed by an increase up to a maximum before
decreasing for larger fields. Below the Burns temperature, this maximum
of $\alpha$ is demonstrated to be correlated to a very specific microscopic
feature, namely to the largest number of dipoles being oriented along
\textbf{$\left\langle 111\right\rangle $} directions having positive
z-component. Finally, equalities that are derived from a simple Landau
model (including one relating $\alpha$ with the product of temperature
and the partial derivative of the square of polarization) reproduce
and further help to understand the anomalous behavior of $\alpha$
with field and temperature in BZT, for any temperature above $T_{m}$
(note that we also found that this model can predict EC effects in
typical ferroelectrics, such as BaTiO$_{3}$, as shown in the Supplemental
Material). Our simulations also confirm that indirect and direct approaches
yield similar results of the $\alpha$ EC coefficient for any temperature
above the $T_{m}$ temperature but differ from each other for temperature
below $T_{m}$, because of the non-ergodicity adopted by BZT at these
low temperatures \cite{Zhang2010,Zhang2014}.

We therefore hope that our study leads to a broader knowledge of EC
effects and relaxor ferroelectrics. 
\begin{acknowledgments}
Z.J., Sergei P. and L.B. thank the DARPA grant HR0011-15-2-0038 (MATRIX
program). Z.J. also acknowledges support from the National Natural
Science Foundation of China (NSFC), Grant No. 51390472, 11574246,
U1537210, National Basic Research Program of China, Grant No. 2015CB654903,
and China Scholarship Council. Sergey P. thanks ONR Grant N00014-12-1-1034,
the grants 3.1649.2017/4.6 from RMES (Russian Ministry of Education
and Science) and 16-52-0072 Bel\_a from RFBR (Russian Foundation for
Basic Research). Y.N. thanks ARO grant W911NF-16-1-0227. We also acknowledge
funding from the Luxembourg National Research Fund through the inter-mobility
(Grant 15/9890527 Greenox, J.I and L.B.) and Pearl (Grant P12/4853155
Cofermat, J.I. and E.D.) programs. Some computations were also made
possible thanks to the MRI grant 0722625 from NSF, the ONR grant N00014-15-1-2881
(DURIP) and a Challenge grant from the Department of Defense.

$^{*}$ \textit{The first two authors contributed equally to this
work}. 
\end{acknowledgments}


\begin{thebibliography}{10}
\bibitem{footnoteequation7} Note that the second line of Eq. (9)
can also be derived from the equality $dT=\frac{2A}{C}TPdP$ provided
in Ref. \cite{Strukov1967}, where $A$ is constant and where $C$
is a specific heat.
\bibitem{squareP} Assuming that $P=\chi{\cal E}$, where $\chi$
is the dielectric susceptibility, yields $\left.\frac{\partial P^{2}}{\partial{\cal E}}\right|_{T}=2\chi^{2}{\cal E}$.
This latter equality therefore explains why $\left.\frac{\partial P^{2}}{\partial{\cal E}}\right|_{T}$
is small for low electric fields, at any temperature.
\end{thebibliography}
\end{document}